\documentclass[twocolumn,noshowpacs,preprintnumbers,amsmath,amssymb]{revtex4}

\usepackage{graphicx}
\usepackage{dcolumn}
\usepackage{bm}
\usepackage{color}

\begin {document}

\title{Capacitively Enhanced Thermal Escape in Underdamped Josephson Junctions}

\author{Y. Yoon,$^{1}$ S. Gasparinetti,$^{1}$ M. M\"{o}tt\"{o}nen,$^{1,2}$ and J. P. Pekola$^{1}$}
\affiliation{$^{1}$Low Temperature Laboratory, Aalto University, P.O.Box 13500, FI-00076 AALTO, Finland \\ $^{2}$Department of Applied Physics/COMP, Aalto University, P.O.Box 14100, FI-00076 AALTO, Finland}

\begin {abstract}

We have studied experimentally the escape dynamics in underdamped capacitively shunted and unshunted Josephson junctions with submicroampere critical currents below 0.5 K temperatures. In the shunted junctions, thermal activation process was preserved up to the highest temperature where the escape in the unshunted junctions exhibits the phase diffusion. Our observations in the shunted junctions are in good agreement with the standard thermal activation escape, unlike the results in the unshunted junctions.

\end {abstract}
\maketitle

Underdamped Josephson junctions are routinely used as threshold current detectors to readout quantum information in a superconducting circuit by measuring switching events from the superconducting to the normal state (see, e.g. \cite {Vio2002}). In addition, this technique has been used to identify between geometric and dynamic phases in a superconducting charge pump \cite{Mik2008}, and also to measure collective quantum phase slips in the ground state of a Josephson junction chain \cite{Pop2010}. The detector sensitivity is important in the measurement. One can improve it by reducing the critical current of the junction, but then the system typically enters the phase diffusion regime, which is not desirable since in this context one usually wants to fix the phase difference of the superconducting circuit. In this letter, we present a technique to prevent the phase diffusion by adding a large shunt capacitance with the help of atomic-layer deposition (ALD).

The Josephson junction dynamics is well described by a model of a phase particle in a tilted washboard potential \cite {Tin1996}. In this model, as the bias current $I$ is increased toward the critical value $I_C$, the barrier height $\Delta U$ is lowered and the particle, oscillating with the plasma (angular) frequency $\omega_p=(2eI_C/C_J \hbar)^{1/2}$ in a potential well, where $C_J$ is the junction capacitance, can escape out of the well by either thermal activation (TA) or macroscopic quantum tunneling (MQT) process. The crossover between MQT and TA regimes occurs at the temperature $T_{\textrm{cr}}\simeq\hbar\omega_p/(2\pi k_B)$ \cite{Kiv2005}. Above $T_{\textrm{cr}}$, the dominant escape mechanism is TA. The junction behavior also depends on the magnitude of the quality factor $Q$ of the junction at its plasma frequency. In the case of junctions with $Q\ll1$, called overdamped, the escape dynamics occurs over a dissipation barrier \cite{Vio1996}, which is undesirable for a threshold current detector. For $Q\gg1$, so-called underdamped case, the phase particle moves down \emph{freely} or \emph{diffusively} from one well to another depending on the effect of dissipation. When dissipation is weak, the free running particle is still governed by TA process, while when dissipation is strong, the escape dynamics enters the phase diffusion regime \cite {Kiv2005,Kra2005,Man2005,Kra2007}. The appearance of phase diffusion in a hysteretic junction was studied extensively two decades ago by Martinis and Kautz \cite {Mar1989,Kau1990}. How can one avoid the underdamped phase diffusion regime and maintain TA process at high temperatures? According to the phase diagram in Ref. \cite{Kiv2005}, decreasing $T_{\textrm{cr}}$ can help the junction escape from the phase diffusion regime by enhancing TA. This idea can be realized by either decreasing $I_C$ or increasing $C_J$. Among these two solutions, however, only the second one is applicable, since the junction with small $I_C$ has also a low Josephson coupling energy $E_J$, which lowers $\Delta U$. Low $\Delta U$ implies a small tilt of the potential at the point of escape, hence the particle gains a small amount of energy when escaping and can get trapped in the next well instead of running down the potential. Accordingly, $C_J$ is the only parameter remaining for decreasing $T_{\textrm{cr}}$. For a typical junction, the capacitance per unit area is 45 fF/$\mu$m$^{2}$. With this value, although one can implement a large $C_J$ by increasing the junction area, it becomes impractical for the detector junction, since $I_C$ is also proportional to the area. Alternatively, a large shunt capacitance can be added in parallel to $C_J$.

Motivated by this idea, we implemented a shunt capacitance $C_{\textrm{sh}}$ by growing Al$_{2}$O$_{3}$ layer of 5.5 nm thickness $d$ using ALD on top of an Al ground plane of 40 nm thickness. Using electron beam lithography and shadow angle evaporation \cite {Dol1977}, Al/AlO$_{x}$/Al Josephson junctions were fabricated on the ALD oxide. The insulating AlO$_{x}$ layers for the tunnel junctions were formed by standard room temperature oxidation between the evaporation steps of the two Al layers, which allows us to manipulate the junction resistance and $I_C$ by varying the oxidation pressure. For large capacitance, two rectangles with 30$\times$15 $\mu$m$^{2}$ area were defined in the leads near the junctions, overlapping with the ground plane, as shown in Fig. 1(b). The total overlap area $A_\textrm{tot}$ between the two parts connected to the junction on the top and the ground plane at the bottom is roughly 500 $\mu$m$^{2}$ each. As a result, the Josephson junction is capacitively shunted by the parallel plate capacitor which consists of two shunt capacitors $C_0$ in series and on each side of the junction. The total shunt capacitance is $C_\textrm{sh}=C_0/2$ where $C_0 = \epsilon A_\textrm{tot}/d$ and $\epsilon$ is the permittivity of Al$_{2}$O$_{3}$. A similar technique for adding shunt capacitance to proximity Josephson junctions was reported in Refs.\cite {Kra2005, Kra2007}, where the capacitor changes the damping strength in superconductor--two-dimensional electron gas--superconductor Josephson junctions. The leakage resistance from the junction to ground plane was larger than 10 G$\Omega$. Both the unshunted and shunted junctions were fabricated at the same time on the same chip. Note that the total capacitance $C_\textrm{tot}$ is $C_\textrm{tot} \simeq C_J$ in the unshunted junction and $C_\textrm{tot} \simeq C_J+C_{\textrm{sh}}$ in the shunted junction, as shown in Fig.1 (c) and Fig. 1 (d), respectively. All measured data were taken in a $^3$He--$^4$He dilution refrigerator with a base temperature of 50 mK.

\begin{figure}
\begin{center}
\includegraphics {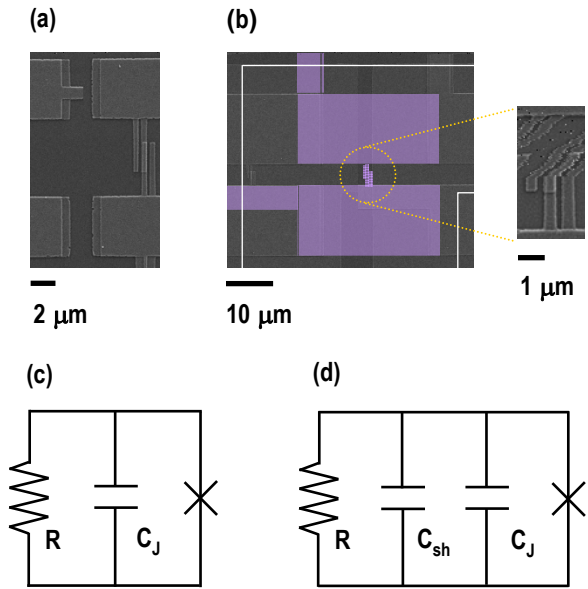} \\
\end{center}

    \caption{\label{fig:1} Scanning electron micrographs of (a) an unshunted and (b) a capacitively shunted junction. In (b), the junction and leads (purple, in foreground) overlap with Al ground plane (encircled by solid white line, in background). The sizes of the junctions in the panel (a) and (b) are 0.4 $\times$ 2.5 $\mu$m$^{2}$ and 0.4 $\times$ 1.5 $\mu$m$^{2}$, and the circuit schematics of the junctions correspond to (c) and (d), respectively. }
\end{figure}

\begin {table} [h]
    \caption {Parameters of the samples. $R_n$ is the normal-state resistance, $I_C$ is the critical current, and $E_J$ is the Josephson energy obtained from Ambegaokar--Baratoff formula \cite{Amb1963}. $C_{\textrm{tot}}$ is the total capacitance including the junction capacitance $C_J$ and the shunted capacitance $C_{\textrm{sh}}$. }
    \centering
\begin {tabular} {c c c c c c c}
\\
\hline\hline\noalign{\smallskip}
&&$R_n$ &$I_C$  &$E_J$ &$C_{\textrm{tot}}$ \\
\raisebox{1.5ex} {Sample} & \raisebox{1.5ex} {Device} &($\Omega$) &(nA) &(K) &(pF)
\\[1ex] \hline \noalign{\smallskip}
 & Unshunted (A1) & 612  & 513 & 12.2 & 0.045  \\ [-1ex]
\raisebox{1.5ex} {A} & Shunted (A2) & 828  & 379 & 9.0 & 3.2  \\ [1ex] \hline \noalign{\smallskip}
 & Unshunted (B1) & 775  & 405 & 9.7 & 0.045 \\ [-1ex]
\raisebox{1.5ex} {B}  & Shunted (B2) & 746  & 421 & 10.0 & 3.2 \\ [1ex]
\hline\hline

\end {tabular}
\end {table}

In Table I, we present the characteristics of the devices studied here. The normal-state resistance $R_n$ was obtained from current--voltage (IV) measurement, which yields $I_C$ and $E_J$ according to Ambegaokar--Baratoff formula \cite{Amb1963}. By the standard parallel plate capacitance calculation based upon the information above and the relative dielectric constant 7.8 of Al$_{2}$O$_{3}$ in ALD processing, $C_{\textrm{sh}}$ was estimated to be 3.14 pF. We also estimated $C_J$ from the specific capacitance of 45 fF/($\mu$m$)^{2}$ and the junction areas given in Figs. 1(a) and (b). Consequently, the total capacitances of the shunted and unshunted junction were 3.2 pF and 45 fF, respectively.

\begin{figure}
\begin{center}
\includegraphics {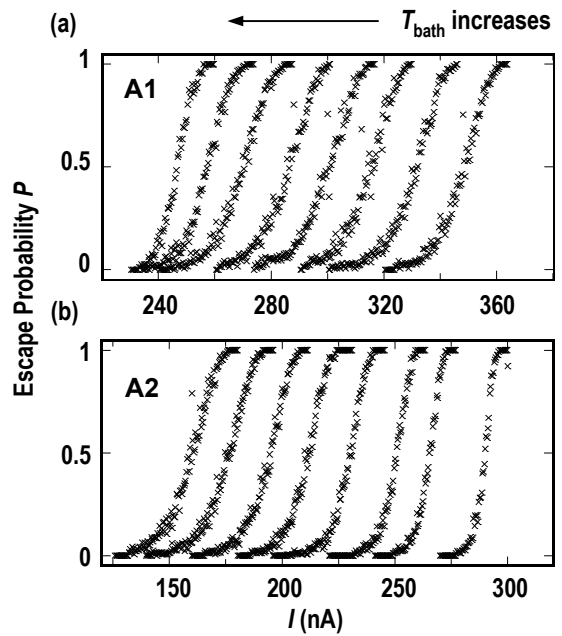} \\
\end{center}

\caption{\label{fig:2} {Measured switching histograms of (a) the unshunted junction A1 and (b) the shunted junction A2. The bath temperature $T_\textrm{bath}$ was set to 58, 120, 160, 216, 270, 321, 377, and 430 mK for the different sets of data from the right to the left. }}

\end{figure}

\begin{figure}
\begin{center}
\includegraphics {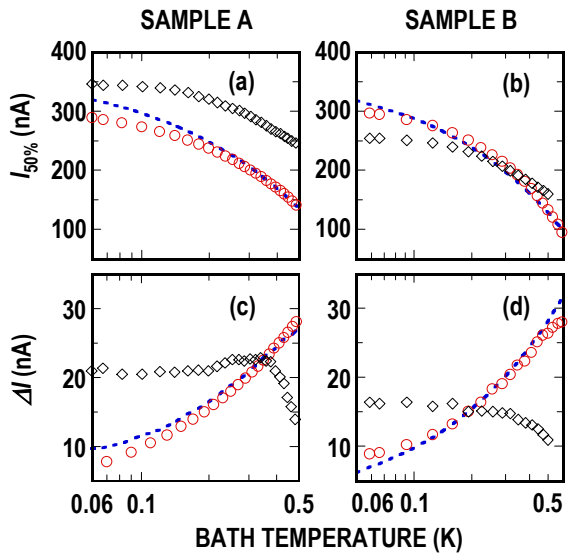} \\
\end{center}

{\caption{\label{fig:3} {The temperature dependent variation of the mean position ($I_{50\%}$) in (a)--(b) and the width ($\Delta I$) of the histograms in (c)--(d). The experimental data in panels (a) and (c) and in (b) and (d) were obtained from Sample A and Sample B, respectively. In each graph, diamond ($\diamond$),  circle ($\circ$), and dashed line correspond to the experimental data of the unshunted junction (A1 and B1), the experimental data of the shunted junction (A2 and B2), and the theoretical simulation of the shunted junction by the TA model, respectively. The simulation was performed with $I_C$ of 379 nA for Sample A and with 369 nA for Sample B as fit parameters, and $C_\textrm{tot}$ = 3.17 pF was employed. The latter $I_C$ differs by 12 \% from the Ambegaokar--Baratoff value. In the TA model, $\Delta I$ does not depend on $C_\textrm{tot}$, or at least the dependence should be very week.}}}

\end{figure}

We investigate the escape dynamics by applying a set of current pulses with a fixed amplitude and duration through the junctions and by determining the probability for the devices to switch from the zero-voltage state to the finite-voltage state. At each amplitude 200 current pulses were applied. The escape probability $P$ was measured as a function of the current pulse amplitude $I$, which yields cumulative histograms of switching current. Figure 2 shows the measured switching histograms of both the unshunted (A1) and the shunted junctions (A2) in Sample A at different bath temperatures. A similar measurement was carried out with Sample B, but the histogram curves are not shown here. The histogram is typically characterized by the switching position $I_{50\%}=I$($P=0.5$), and the width $\Delta I$ = $I(P=0.9)$ $-$ $I (P=0.1)$, the measurement results of which are shown in Fig. 3. Note that at the lowest temperature, the switching position of A2 is lower than that of A1, as shown in Fig. 3(a), and Sample B behaves in an opposite manner as shown in Fig. 3(b). We observed that on increasing bath temperature $T_\textrm{bath}$ the mean position shifts down in all cases, whereas the temperature dependence of the width is very different depending on whether the junction is capacitively shunted to the ground plane or not. According to the standard MQT and TA discussed above, $T_{\textrm{cr}}$ $\approx$ 11 mK $\ll T_\textrm{bath}$ in the shunted junctions (A2 and B2). Thus the role of the MQT is negligible and instead, TA dominates the escape dynamics. In the TA model, the width obeys $\Delta I$ $\propto$ $T_\textrm{bath}^{2/3}$. Consistent with the prediction of the model, we observed that in the shunted junctions the widths are increasing as a function of temperature. The dashed lines in Fig. 3 show that the theory is in quantitative agreement with our observations. For the corresponding unshunted junctions (A1 and B1), on the other hand, on increasing $T_\textrm{bath}$ the widths are more or less constant up to $T_{\textrm{cr}}$ $\approx$ 200 mK, as predicted with the MQT model, but they decrease at high temperatures. It was reported that the decreasing $\Delta I$ results from the influence of dissipation by a retrapping process \cite {Kiv2005, Man2005, Kra2005, Kra2007}. This indicates phase diffusion. The TA model cannot explain the phenomenon because it neglects the effect of dissipation. Several models have been proposed to account for the phase diffusion of underdamped Josephson junctions incorporating the retrapping process in the TA model \cite {Kiv2005, Man2005, Kra2005, Kra2007, Fen2008}.

Kivioja \emph{et al.} \cite {Kiv2005} claim that when $I$ is below a maximum possible phase diffusion current $I_m$ = 4$I_C$/$\pi$$Q$, the phase particle is retrapped after escape. The observed $I_m$ is 265 nA in A1 and 190 nA in B1, obtained from $I_{50\%}$ curves at a certain temperature where $\Delta I$ starts to decrease in Fig. 3. With those values and $I_C$ given in Table I, $Q$ of the unshunted junctions is estimated to be 2.5 in A1 in and 2.7 in B1. Note that this $Q$ may differ from that introduced earlier since here it corresponds to low frequency dynamics of the junction. For the shunted junctions, however, it is not simple to determine $Q$ with the present information. Nevertheless, we believe that $Q$ must be at least an order of magnitude larger than those of the unshunted junctions, because of the large $C_\textrm{tot}$.

In conclusion, we developed a technique to add a large shunt capacitance using atomic-layer deposition in a Josephson junction to prevent the junction from entering phase diffusion. We compared the escape dynamics of the capacitively shunted and unshunted junctions. In the shunted junction, TA process is preserved at all measured temperatures; for the unshunted junctions we observe phase diffusion. The TA model yields a satisfactory discussion for our observations in the shunted junctions. It may be possible that, by engineering this shunt capacitance for the experiments on the phase biased Cooper pair pumps \cite {Mik2008}, we can not only improve the sensitivity of the detector junction but also protect the circuit from high frequency noise. However, attention has to be paid on the fluctuations in the phase bias due to thermal effects.

\begin{acknowledgements}
The authors gratefully acknowledge discussions with P. Solinas, M. Meschke, and J. Peltonen. This work was supported by the European Community's Seventh Framework Programme under Grant No. 238345 (GEOMDISS).
\end{acknowledgements}

\end {document}